\documentstyle[stwol,amstex,amssymb]{article}
\input{psfig}
\title{\uppercase{Thermal Flavor Production and\\ Signatures of 
Deconfinement
}}
\author{\uppercase {J. Rafelski} }

\address{Department of Physics, University of Arizona, Tucson, AZ
85721, USA}

\author{\uppercase {J.  Letessier \lowercase {and} A. Tounsi}}

\address{Laboratoire de Physique Th\'eorique et Hautes Energies\\
Universit\'e Paris 7, Tour 24, 2 Pl. Jussieu, F-75251
Cedex 05, France}

\begin{document}

\twocolumn[\maketitle\abstracts
{
\vspace*{-6cm}
\centerline{Published in Proceeding of ICHEP'96, Warsaw, p.\,971, Z. Ajduk and K. Wrobleski, Eds., World Scientific, Singapore (1997).}
\vspace*{6cm}
Using  renormalization group methods we evaluate the thermal strangeness 
and charm chemical equilibrium relaxation times in the deconfined 
quark-gluon plasma.  We present a reaction model and evaluate the total 
production rate of strangeness  in fixed target
Pb--Pb collisions at 10--300 A GeV. 
We discuss the relevance of our results to the diagnosis 
and understanding of the properties of the deconfined state.}]

\section{Introduction}
In high energy nuclear collisions\,\cite{Sto97}
 we are interested to create and study the 
deconfined quark-gluon phase, believed to consist in the limit of very high 
energy densities of a liquid of quarks and gluons interacting perturbatively.
We are motivated by the desire to recreate in the laboratory conditions akin
to those prevailing in the first moments of the Early Universe, and in the 
recognition that we can study the properties of the excited, `melted', 
vacuum state of strong interactions.

Since in the collision of large  nuclei  the highly dense state is 
formed for a rather short time\,\cite{Bjo83}\!, one of the major challenges 
has been to identify suitable physical observables of deconfinement. 
A number of possible experimental signatures of the formation 
and properties of quark-gluon plasma (QGP) have been
studied. This report addresses our recent advances in evaluating 
the relevance of strange\,\cite{Ody97,acta96} and charm quark flavor 
in this quest\,\cite{acta96}. 
Other major probes include the  phenomenon of $J\!/\!\Psi$ suppression 
\cite{Pet97}, photons  and  dileptons\,\cite{Pfe97,Red97}. 

Strangeness is a very interesting diagnostic tool of dense hadronic 
matter\,\cite{S95}:\\
\indent {\bf 1)} particles containing  strangeness are found more 
abundantly in relativistic nuclear collisions than it could be 
expected based on simple scaling of $p$--$p$ reactions;\\ 
\indent {\bf 2)} all strange hadrons have to be
made in inelastic reactions, while light $u$, $d$  quarks are  also
brought into the reaction by the colliding nuclei;\\ 
\indent {\bf 3)}
because there are many different strange particles, we have a
very rich field of observables with which it is possible to explore
diverse properties of the source;\\ 
\indent {\bf 4)} theoretical calculations suggest that glue--glue
collisions in the QGP provide a sufficiently fast and
thus by far a unique mechanism leading to an explanation of
strangeness enhancement. 
 
We begin by recalling these mechanisms of strangeness production. 
We then obtain in section \ref{runqcd} the magnitude of 
running coupling constant  and quark  masses.
Using these results we evaluate the thermal relaxation times of
strangeness and charm as function  of temperature 
in section \ref{thermrat}.
We discuss the relevance of the flavor observable of QGP and
describe how these relaxation times allow to compute the hadronic
particle yields in  section \ref{gammas}. We close with a few general 
remarks about the relevance of our results.

\section{Strangeness Production in QGP} \label{sQGP}
We will now show how to evaluate using two particle collision 
processes flavor 
production\,\cite{sprodQGP,impact96} in thermal QGP. Ultimately, 
we will employ  running QCD parameters $\alpha_{\rm s}(\mu)$ 
and $m_i(\mu),\,i=$s,\,c. While theory and experiment constrain now
sufficiently the coupling strength $\alpha_{\rm s}$, considerable
uncertainty still remains in particular in regard of strange quark
mass scale, as well as systematic uncertainty related to applications
of QCD to soft (less than 1 GeV) processes.

The generic angle averaged two particle cross section for (heavy)
flavor production processes
\mbox{$g+g\to f+\bar f $ and $ q+\bar q\to f+\bar f\,,$ are:}
\begin{align}
\bar\sigma_{gg\to f\bar f}(s) &=
   {2\pi\alpha_{\rm s}^2\over 3s} \left[
\left( 1 + {4m_{\rm f}^2\over s} + {m_{\rm f}^4\over s^2} \right)
\right.\label{gl}\\
&\hspace*{-0.02cm}\left.\cdot\    
{\rm tanh}^{-1}W(s)-\left({7\over 8} + {31m_{\rm f}^2\over
8s}\right) W(s) \right],\nonumber\\
\bar\sigma_{q\bar q\to f\bar f}(s) &=
   {8\pi\alpha_{\rm s}^2\over 27s}
   \left(1+ {2m_{\rm f}^2\over s} \right) W(s)\,,
\label{gk}
\end{align}
where $W(s) = \sqrt{1 - 4m_{\rm f}^2/s}$\,, and both the QCD
coupling constant $\alpha_{\rm s}$  and flavor quark mass $m_{\rm
f}$ will be in this work the running QCD parameters. 
In this way a large number of even-$\alpha_{\rm s}$
diagrams contributing to flavor production  is accounted for.
 
What remains unaccounted for is another class of processes 
in which at least one additional gluon is present. In
particular processes allowing the production of
an additional soft gluon in the final state remains unaccounted for
today. Leading diagrams contain odd powers of $\alpha_{\rm s}$ and
their generic cross section is in general infrared divergent,
requiring a cut-off which for processes occurring in matter is
provided by the interactions (dressing) with other particles
present. The process in which a massive `gluon', that is a
quasi-particle with quantum numbers of a gluon, decays into a
strange quark pair, is partially included in the resummation that
we accomplish in the present work. At the present time we do not
see a systematic way to incorporate any residue of this and other
effects, originating in matter surrounding the microscopic
processes, as work leading to understanding of renormalization
group equations in matter (that is at finite temperature and/or
chemical potential) is still in progress\,\cite{Elm95}. 
 
\section{Running QCD Parameters} \label{runqcd}
To determine the two QCD parameters required, we will use the
renormalization group functions $\beta$ and $\gamma_{\rm m}$:
\begin{equation}\label{dmuda}  
\hspace*{-0.5cm}\mu \frac{\partial\alpha_{\rm s}}{\partial\mu}
=\beta(\alpha_{\rm s}(\mu))\,,\
\mu {\frac{\partial m}{\partial\mu}} =-m\,
\gamma_{\rm m}(\alpha_{\rm s}(\mu))\,.\hspace*{-0.5cm}
\end{equation}
For our present study we will use the perturbative power expansion 
in $\alpha_{\rm s} $:
\begin{align}\label{betaf}
\beta^{\rm pert}&=\alpha_{\rm s}^2\left[\ b_0
   +b_1\alpha_{\rm s} +b_2\alpha_{\rm s}^2 +\ldots\ \right] \,,\nonumber\\
\label{gamrun}
\gamma_{\rm m}^{\rm pert}&=\alpha_{\rm s}\left[\ c_0
+c_1\alpha_{\rm s} +c_2\alpha_{\rm s}^2 + \ldots\ \right]\,,
\end{align}
For the SU(3)-gauge theory with $n_{\rm f}$ fermions the first two
terms (two `loop' order) are renormalization scheme independent, and
we include in our calculations the three `loop' term as well, which
is  renormalization scheme dependent, evaluated in the 
MS-scheme\,\cite{SS96}. We have:
\begin{align}
b_0\!=&\,\displaystyle\frac{1}{ 2\pi}
  \left(\!11\!-\!{2\over 3}n_{\rm f}\!\right)\!,\hspace*{0.3cm}
b_1\!=\,\frac{1}{4\pi^2}\left(\!51\!-\!{19\over 3}
        n_{\rm f}\!\right), \hspace*{-0.3cm}\\
b_2\!=&\,\displaystyle\frac{1}{64\pi^3}\left(\!2857-{\frac {5033}{9}}\,n_{\rm f}
+{\frac {325}{27}}\,{n_{\rm f}}^{2}\!\right),\ \nonumber\\
c_0\!=&\, \displaystyle\frac{2}{\pi},\hspace*{0.3cm}  
c_1\!=\frac{1}{12\pi^2}
        \left(\!101\!-\!{10\over 3}n_{\rm f}\!\right),\\
c_2\!=&\,\displaystyle\frac{1}{32\pi^3}\!\left(\!1249
\!-\!\left(\!{2216\over 27}\!+\!{160\over 3}\zeta(3)\!\right)\!n_{\rm f}
\!-\!{140\over 81}{n_{\rm f}}^{2}\!\right)\!.\nonumber
\end{align}
The number  $n_{\rm f}$ of fermions that can be excited, depends
on the energy scale $\mu$. We have implemented this using the
exact phase space form appropriate for the terms linear
in $n_{\rm f}$:
\begin{align}\label{nfs}
n_{\rm f}(\mu)=2+\sum_{i=s,c,b,t}&\sqrt{1-\frac{4m_i^2}{\mu^2}}\\
  &\cdot\ \left(1+\frac{2m_i^2}{\mu}\right)\Theta(\mu-2m_i)\,,\nonumber
\end{align}
with $m_{\rm s}\!=\!0.16\,{\rm GeV},\,m_{\rm c}\!=\!1.5\,{\rm GeV},\,
m_b\!=\!4.8$\,{\rm GeV}. We checked that there is very minimal impact
of the running of the masses in Eq.\,(\ref{nfs}) on the final
result, and will therefore not introduce that `feed-back' effect
into our current discussion. The largest effect on our solutions
comes from the bottom mass, since any error made at about 5 GeV is
amplified most. However, we find that this results in a scarcely
visible change even when the mass is changed by 10\% and thus 
one can conclude that the exact values of the masses and the 
nature of flavor threshold is at present of minor importance in our
study. 
 
\begin{figure}[!t]
\vspace*{-.8cm}
\centerline{\hspace*{0.3cm}
\psfig{width=10cm,figure=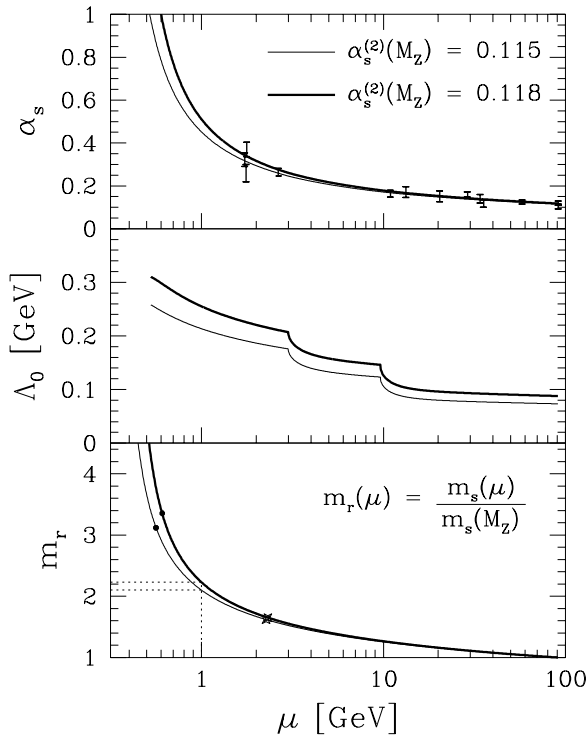}
}
\vspace*{0.2cm}
\caption{
\rm$\alpha_{\rm s}(\mu)$ (top section); the equivalent 
parameter $\Lambda_0$ (middle section) and
$m_{\rm r}(\mu)=m(\mu)/m(M_{Z})$
(bottom section) as function of energy scale $\mu$. 
Initial value $\alpha_{\rm s}(M_{Z})=0.118$ (thick solid lines)
and $\alpha_{\rm s}(M_{Z})=0.115$ (thin solid lines).
In lower section the dots indicate the strangeness pair production
thresholds for $m_{\rm s}(M_{Z})=$~90~MeV, while crosses indicate
charm pair production thresholds for $m_{\rm c}(M_{Z})=$~700~MeV.
\label{fig-a1}}
\end{figure}
We show the result of numerical integration for $\alpha_{\rm s}$ in
the top portion of Fig.\,\ref{fig-a1}. First equation in
(\ref{dmuda}) is numerically integrated  beginning with an initial
value of $\alpha_{\rm s}(M_Z)$. We use in this report
the August 1996 World average\,\cite{ICHEP96}: 
$\alpha_{\rm s}(M_{{Z}})=0.118$ for which the estimated error is 
$\pm\ 0.003$\,. This value is sufficiently precise to eliminate most
of the uncertainty that has befallen much of our earlier
studies\,\cite{acta96,impact96}. In addition, the thin solid lines
present results for $\alpha_{\rm s}(M_{{Z}})=0.115$\, till recently
the preferred result in some analysis, especially those at lower
energy scale. As seen in Fig.\,\ref{fig-a1}, the variation of
$\alpha_{\rm s}$ with the energy scale is substantial, and in
particular we note the rapid change at and below $\mu=1$ GeV, where
the strange quark flavor formation occurs in hot QGP phase formed
in present day experiments at 160--200 A GeV (SPS-CERN). Clearly,
use of constant value of $\alpha_{\rm s}$ is hardly justified, and
the first order approximation often used:
\begin{equation}\label{Lambdarun}
\alpha_{\rm s}(\mu)\equiv
  \frac{2b_0^{-1}(n_{\rm f})}{\ln(\mu/\Lambda_0(\mu))^2}\,,
\end{equation}
leads to a strongly scale dependent $\Lambda_0(\mu)$ shown in the
middle  section of  Fig.\,\ref{fig-a1}. Thus it also cannot be used
in the evaluation of thermal strangeness and charm production.

With $\alpha_{\rm s}(\mu)$ from the solutions described above, we
integrate the running of the quark masses, the second  equation in
(\ref{dmuda}).  Because the running mass equation 
is linear in $m$, it is possible to determine the universal 
quark mass scale factor
\begin{equation}
m_{\rm r}=m(\mu)/m(\mu_0)\,.
\end{equation}
Since  $\alpha_{\rm s}$ refers to  the scale of $\mu_0=
M_Z$, it is a convenient reference point also for quark masses.  
As seen in the bottom portion of Fig.\,\ref{fig-a1},
the change in the quark mass factor is highly relevant,
since it is driven by the rapidly changing
$\alpha_{\rm s}$ near to $\mu\simeq 1$~GeV.
For each of the different functional dependences
$\alpha_{\rm s}(\mu)$ we obtain a different function
$m_{\rm r}$. The significance of the running of the charmed
quark mass cannot be stressed enough, especially for thermal charm
production occurring in foreseeable future experiments well below
threshold, which amplifies the importance of exact value of 
$m_{\rm c}$\,.
 
Given these results, we find that for $\alpha_{\rm s}=0.118$ and
$m_{\rm s}(M_{{Z}})=90\pm18$~MeV a low energy strange quark mass
$m_{\rm s}(1\mbox{\,GeV})\simeq 200\pm 40$ MeV, in the middle of
the standard range $100<m_s$(1\,GeV) $<$ 300 MeV. Similarly we
consider $m_{\rm c}(M_{{Z}})=700\pm50$~MeV, for which value 
we find  the low energy mass $m_c(1\mbox{\,GeV})\simeq 1550\pm110$
MeV, at the upper (conservative for particle production yield) end
of the standard range $1<m_c$(1\,GeV) $<1.6$ MeV. There is another
(nonperturbative) effect of mass running, related to the mass at
threshold for pair production $m^{\rm th}_i,\,i=$ s, c, arising from 
the solution of:
\begin{equation}\label{dispersion} 
m_i^{\rm th}/m_i(M_{{Z}})= m_{\rm r}(2m_i^{\rm th})\,.
\end{equation}
This effect stabilizes strangeness production cross section in the
infrared: below $\sqrt{\rm s}=1$ GeV the strange quark mass
increases rapidly and the threshold mass 
is considerably greater than  $m_{\rm s}$(1 GeV).
We obtain the threshold values $2m_{\rm s}^{\rm th}=611$ MeV
for  $\alpha_{\rm s}(M_Z)=0.118$ 
and $2m_{\rm s}^{\rm th}=566$ MeV for $\alpha_{\rm s}(M_Z)=0.115$. 
Both values are indicated by the black dots in 
Fig.\,\ref{fig-a1}.  For charm, the running mass effect plays
differently:  since the mass of charmed quarks is listed in tables
for $\mu=1$ GeV, but the value of the  mass is above 1 GeV,  the
production threshold mass is smaller than expected (i.e., listed
value). For $m_{\rm c}(M_Z)= 700$ MeV the
production threshold is found at $\sim 2m_c^{\rm th}\simeq 2.3$ GeV
rather than 3.1 GeV that would have been expected for the $m_{\rm
c}$(1 GeV). This reduction in threshold enhances thermal production
of charm, especially so at low temperatures.
\nopagebreak 
\section{Strangeness and Charm Thermal Relaxation Times}
\label{thermrat}
\begin{figure}[!t]
\vspace*{-0.3cm}
\centerline{\hspace*{-0.5cm}
\psfig{width=8.5cm,figure=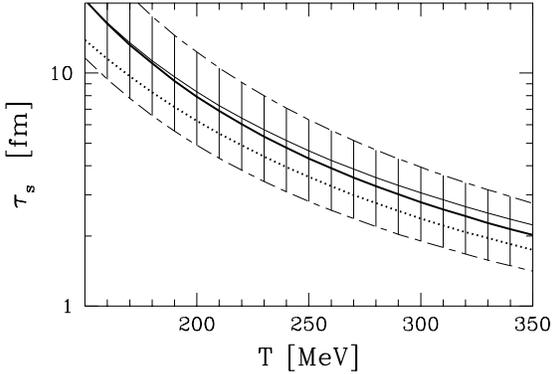}}
\vspace*{-0.3cm}
\caption{
QGP strangeness relaxation time, for $\alpha_{\rm s}(M_{Z})=0.118$,
(thick line) and = 0.115 (thin line); $m_{\rm s}(M_{{Z}})=90$~MeV.
Hatched areas: effect of variation of strange quark mass by 20\%.
 Dotted: comparison results for fixed  
$\alpha_{\rm s}=0.5$ and $m_{\rm s}=200$ MeV. 
}
\label{figTaussrun}
\end{figure}
The thermal average of the cross section is the invariant
production rate per unit time and volume:
\begin{align}
A_{\rm s}\equiv&\, A_{gg}+A_{u\bar u}+A_{d\bar
d}+\ldots\nonumber\\
=&\int_{4m_{\rm s}^2}^{\infty}ds
2s\delta (s-(p_1+p_2)^2)
\int{d^3p_1\over2(2\pi)^3E_1}\nonumber\\
&\times\int{d^3p_2\over2(2\pi)^3 E_2}
\left[{1\over 2} g_g^2f_g(p_1)f_g(p_2)
\overline{\sigma_{gg}}(s)\right.\nonumber\\ 
&\hspace*{0.5cm}\left.+ n_{\rm f}g_q^2 f_q(p_1) 
f_{\bar q}(p_2)\overline{\sigma_{q\bar
q}}(s)+\ldots\vphantom{{1\over 2}}\right]\!.\hspace*{-0.5cm}\label{qgpA} 
\end{align}
The dots indicate that other mechanisms may contribute to
strangeness production. The particle distributions $f_i$ are in our
case thermal Bose/Fermi functions (for fermions with $\lambda_{\rm
q}=1.5$), and $g_{\rm q}=6,\,g_{\rm g}=16$\,. For strangeness
production $n_{\rm f}=2$, and for charm production $n_{\rm f}=3$\,.
>From the invariant rate we obtain the strangeness relaxation time
$\tau_{\rm s}$ shown in Fig.\,\ref{figTaussrun}, as function of
temperature:
\begin{equation}\label{tauss}
\tau_{\rm s}\equiv
{1\over 2}{\rho_{\rm s}^\infty(\tilde m_{\rm
s})\over{(A_{gg}+A_{qq}+\ldots)}}\,. 
\end{equation}
Note that here unaccounted for processes, such as the above
mentioned odd-order in $\alpha_{\rm s}$ would add to the production
rate incoherently, since they can be distinguished by the presence
of incoming/outgoing gluons. Thus the current calculation offers an
upper limit on the actual relaxation time, which may still be
smaller. In any case,  the present result suffices to confirm
that strangeness will be very near to chemical equilibrium in QGP
formed in collisions of large nuclei. 
 
We  show in Fig.\,\ref{figTaussrun}  also
the impact of a 20\% uncertainty in $m_{\rm s}(M_{{Z}})$, 
indicated by the hatched areas. This uncertainty is today
much larger compared to the uncertainty that arises from the
recently improved precision of the strong coupling constant
determination\,\cite{ICHEP96}.  We note that
the calculations made\,\cite{sprodQGP} at fixed values 
$\alpha_{\rm s}=0.5$ and $m_{\rm s}=200$~MeV 
(dotted line in Fig.\,\ref{figTaussrun}) are well within the band
of values related to the uncertainty in the strange quark mass.
 
\begin{figure}[!t]
\vspace*{-0.3cm}
\centerline{\hspace*{-0.5cm}
\psfig{width=8.5cm,figure=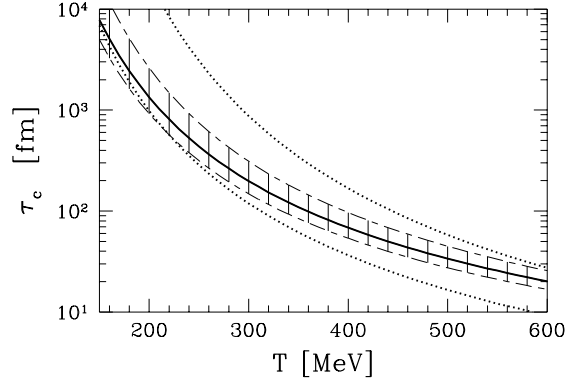}
}
\vspace*{-0.3cm}
\caption{
Solid lines: thermal charm relaxation constant in QGP, calculated
for running  $\alpha_s(M_{Z})=$ 0.115; 0.118, (indistinguishable),
$m_{\rm c}(M_Z)= 700$ MeV. Lower dotted line: for fixed $m_{\rm
c}=1.1$ GeV, $\alpha_{\rm s}=0.35$;
upper doted  line: for fixed $m_{\rm c}=1.5$ GeV, $\alpha_{\rm
s}=0.4$\,.
Hatched area: effect of variation $m_{\rm c}(M_Z)= 700\pm 50$~MeV
\label{figtaucc}
}
\end{figure}

Since charm is somewhat more massive compared to strangeness, there
is still less uncertainty arising in the extrapolation of the
coupling constant. Also the systematic uncertainty related to the
soft gluons (odd-$\alpha_{\rm s}$) terms are smaller, and thus the
relaxation times $\tau_{\rm c}$ we show in Fig.\,\ref{figtaucc}
are considerably better defined compared to $\tau_{\rm s}$. There
is also  less relative uncertainty in the value of charm mass.
We also show in Fig.\,\ref{figtaucc} (dotted lines) the fixed 
$m_{\rm c},\, \alpha_{\rm s}$ results with parameters selected to
border high and low $T$ limits of the results presented. It is
difficult to find a good comparative behavior of $\tau_{\rm c}$ 
using just one set of $m_{\rm c}$ and $\alpha_{\rm s}$. This may be
attributed to the importance of the mass of the charmed quarks,
considering that the threshold for charm production is well above
the average thermal collision energy, which results in emphasis of
the effect of running charm mass. In the high $T$-limit the choice
(upper doted line in Fig.\,\ref{figtaucc}) $m_{\rm c}=1.5$ GeV,
$\alpha_{\rm s}=0.4$ is appropriate, while to follow
the result at small $T$  (lower doted line in
Fig.\,\ref{figTaussrun}) we take a much smaller 
mass $m_{\rm c}=1.1$ GeV, $\alpha_{\rm s}=0.35$\,.
 
We recall that the equilibrium distribution is result of Boltzmann 
equation description of two body collisions. Thus the mass arising
in the equilibrium density $\rho_{\rm s}^\infty$ in
Eq.\,(\ref{tauss}) is to be taken at the energy scale of the
average two parton collision. We adopt for this purpose a fixed
value ${\tilde m}_{\rm s}=200$~MeV, and observe that in the range
of temperatures here considered the precise value of the mass is
insignificant, since the quark density is primarily governed by the
$T^3$ term in this limit, with finite mass correction being ${\cal
O}$(10\%). The situation is less clear for charm relaxation, since
the running of the mass should have a significant impact. Short of
more complete kinetic treatment, we used $m_{\rm c}\simeq 1.5$ GeV
in order to establish the reference density $\rho_{\rm c}^\infty$
in Eq.\,(\ref{tauss}). 
  
\section{QGP Flavor Observable}\label{gammas}
We will indicate in this section how the study of flavor production
impacts  our understanding and diagnosis of the deconfined QGP
phase. We recall first that there are two generic flavor observable
which we can study  analyzing experimental data:\\
\noindent {\bf$\bullet$ yield of strangeness/charm:}\\ 
{\it once produced in hot early QGP phase, strangeness/charm is
not reannihilated in the evolution of the deconfined state towards
freeze-out, and thus the flavor yield is characteristic of the
initial, most extreme conditions;} \\
\mathversion{bold}
\noindent {\bf$\bullet$ phase space occupancy 
$\gamma_{\rm s,c}$:}\mathversion{normal}\\
{\it impacts distribution of flavor among final state particle
abundances.} 

Given that the thermal equilibrium is established within a
considerably shorter time scale than the (absolute) heavy flavor
chemical equilibration, we can characterize the equilibration
of the  phase space occupancy by an average over the momentum
distribution:
\begin{equation}\label{gamth}
\gamma_i(t)\equiv {
           \int d^3\!p\,d^3\!x\,n_i(\vec p,\vec x;t)\over 
     \int d^3\!p\,d^3\!x\,n_i^{\infty}(\vec p,\vec x)}\,,\
i={\rm s,\,c}\,.\hspace*{-0.5cm}
\end{equation}
The chemical equilibrium density is indicated by upper-script
`$\infty$'. When several carriers of the flavor are present, as is
the case in the confined phase, $n_i$ is understood to comprise a
weighted sum.
 
In order to be able to compute the production and evolution of
strangeness and charm flavor a more specific picture of the
temporal evolution of dense matter is needed. Here, we will address
specifically strangeness production in collisions at CERN-SPS, up
to 200 A GeV per nucleon. We  use a simple, qualitative
description, simplified by the assumption that the properties of
the hot, dense matter are constant across the entire volume
(fireball model). We consider radial expansion to be the dominant
factor for the evolution of the fireball properties such as
temperature/energy density and lifetime of the QGP phase. 
Within the fireball model the expansion dynamics 
follows from two assumptions: \\ 
\noindent $\bullet$
the (radial) expansion 
is entropy conserving, thus the volume and temperature satisfy:
\begin{equation}\label{adiaex}
V\cdot T^3=\,{\rm Const.}
\end{equation}
\noindent $\bullet$
the surface flow velocity is given by the 
sound velocity in  a relativistic gas
\begin{equation}
v_{\rm f}=1/\sqrt{3}\,.
\end{equation}

This leads to the explicit forms for the radius of the
fireball and its 
average temperature:
 \begin{equation}
R=R_{\rm in}+{1\over \sqrt{3}}(t-t_{\rm in}),\
\label{T(t)} 
T={T_{\rm in}
\over{1+({t-t_{\rm in}})/\sqrt{3}R_{\rm in}}}.\nonumber
\end{equation}
The initial conditions for Pb--Pb:

\noindent$\lambda_{\rm q}=1.6,\ t_{\rm in}=1\,{\rm fm}/c,
\ T_{\rm in}=320\,{\rm MeV;}$ with

\noindent$R_{\rm in}=4.5\,{\rm fm}$ for $\eta=0.5$; 
$R_{\rm in}=5.2\,{\rm fm}$ for $\eta=0.75$,

\noindent and for S--Pb/W:

\noindent$\lambda_{\rm q}=1.5,\ t_{\rm in}=1\,{\rm fm}/c,\ 
T_{\rm in}=280\,{\rm MeV;}$ with

\noindent$R_{\rm in}=3.3\,{\rm fm}$ for $\eta=0.35$; 
$R_{\rm in}=3.7\,{\rm fm}$ for $\eta=0.5$,

\noindent have been determined such that the energy per baryon is 
given by energy and baryon flow, and the total baryon number is 
$\eta (A_1+A_2)$, as stopped in the interaction region. The radius 
shown above are for zero impact parameter.
For this, equations of state of the 
QGP are needed, and we have employed our model\,\cite{init} in which the
perturbative correction to the number of degrees of freedom were 
incorporated along with thermal particle masses.
 
\begin{table*}[!t]
\caption{$\gamma_{\rm s}$ and  $N_{\rm s}/B$ 
in S--W at 200 A GeV and Pb--Pb at 158 A GeV
for different stopping values of baryonic number and energy 
$\eta_{\rm B}=\eta_{\rm E}$\,; computed for strange quark mass 
$m_s(1 GeV)=200\pm40$ MeV, $\alpha_{\rm s}(M_Z)=0.118$\,.}
\vspace{0.4cm}
\begin{center} 
\begin{tabular}{|c||c|c|c|c|} 
\hline\vphantom{$\displaystyle\frac{E}{B}$}
$E_{\rm lab}$&\multicolumn{2}{|c|}{S--W at 200 A GeV}&%
\multicolumn{2}{|c|}{Pb--Pb at 158 A GeV}\\
\hline\hline\vphantom{$\displaystyle\frac{E}{B}$}
$\eta_{\rm B}=\eta_{\rm E}$&0.35&0.5&0.5&0.75  \\
\hline\vphantom{$\displaystyle\frac{E}{B}$}
$\gamma_{\rm s}$& $0.53\pm 0.14$   &$0.65\pm 0.15$ &%
$0.69\pm 0.15$ &$0.76\pm 0.16$ \\
\hline\vphantom{$\displaystyle\frac{E}{B}$}
$N_{\rm s}/B$&$0.67\pm 0.16$&$0.70\pm 0.16$&%
$0.954\pm 0.20$&$0.950\pm 0.20$\\
\hline
\end{tabular} 
\end{center} 
\end{table*} 

In the fireball in every volume element we have:
\begin{equation}\label{gamdef2}
n_{\rm s}(\vec p;t)=\gamma_{\rm s}n_{\rm s}^\infty
     (\vec p;T,\mu_{\rm s})\,.
\end{equation} 
In this limit and allowing for the detailed balance reactions, thus
re-annihilation of flavor, the yield is obtained from the equation:
\begin{equation}\label{dNsdt}
{{dN_{\rm s}(t)}\over {dt}} = 
V(t)A_{\rm s}\left[1-\gamma_{\rm s}^2(t)\right]\,.
\end{equation}
Allowing for dilution of the phase space density 
in expansion, we derive\,\cite{acta96} from Eq.\,(\ref{dNsdt}) 
an equations describing the change in $\gamma_{\rm s}(t)$:
\begin{equation}\label{dgdtf}
\hspace*{-0.2cm}{{d\gamma_{\rm s}}\over{dt}}\!=\!
\left(\!\gamma_{\rm s}{{\dot T m_{\rm s}}\over T^2}
     {d\over{dx}}\ln x^2K_2(x)\!+\!
{1\over 2\tau_{\rm s}}\left[1-\gamma_{\rm s}^2\right]\!\right).
\end{equation}
Here K$_2$ is a Bessel function and $x=m_{\rm s}/T$. 
Note that even when $1-\gamma_{\rm s}^2<1$ we still can have 
a positive derivative of $\gamma_{\rm s}$, since the first term
on the right hand side of Eq.\,(\ref{dgdtf}) is always positive,
both $\dot T$ and $d/dx(x^2k_2)$ being always negative. This shows
that dilution due to expansion effects in principle can make the
value of $\gamma_{\rm s}$ rise above unity.
 
Given  the relaxation constant $\tau_{\rm s}(T(t))$, these
equations can be integrated numerically, and we can obtain 
for the two currently explored experimental systems the values of the
two observables, $\gamma_{\rm s}$ and $N_{\rm s}/B$), which are given in 
table 1.

For S--Ag collisions at 200 A GeV a
recent evaluation of the specific strangeness yield leads to 
$N_{\rm s}/B\vert_{\rm exp}=0.86\pm0.14$ (see table~4 of
Ref.\,\cite{GR96}). Our earlier analysis\,\cite{acta96}
of the WA85 data yields  $\gamma_{\rm s}=0.75\pm0.1$ 
for S--W interactions. Both these results are in good agreement
with the theoretical result shown 
in the table, favoring the 50\% stopping case for S--W. The 
analysis of the experimental Pb--Pb data is in progress.

\begin{figure}[!b]
\vspace*{-2.5cm}
\centerline{\hspace*{-2.4cm}
\psfig{width=8.2cm,figure=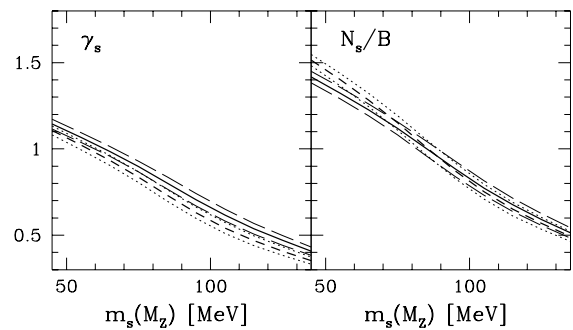}
}
\vspace*{-0.3cm}
\caption{
Phase space occupancy $\gamma_{\rm s}$ and yield of
strange quarks per baryon $N_{\rm s}/B$ as function of strange quark
mass $m_{\rm s}(M_Z)$ for Pb--Pb collision system at 158 A GeV, for the two
stopping fractions and $\alpha_{\rm s}(M_Z)=0.118\pm 0.003$\,:
thick solid lines, $\eta=75\%$\,, with long dashed lines indicating 
the $\alpha_{\rm s}(M_Z)$ uncertainty, thick dashed lines for
$\eta=50\%$\,, with dotted lines indicating 
the $\alpha_{\rm s}(M_Z)$ uncertainty.
\label{fignsbms}
}
\end{figure}
As we can see in table 1, there is a considerable uncertainty due to 
the unknown mass of strange quarks. On the other hand, inspecting 
the yield of strange quarks per baryon $N_{\rm s}/B$ there seem to be 
very little dependence on the stopping fractions. This insensitivity 
to the reaction mechanism coupled to a visible sensitivity 
to strange quark mass suggests that additional insight may be ultimately
gained about the strange quark mass from the study of strangeness enhancement
in relativistic heavy ion collisions. Thus we show in Fig.\,\ref{fignsbms}
as function of $m_{\rm s}(M_Z)$ for the two different stopping
fractions the resulting strangeness yield per baryon, $N_{\rm s}/B$. 
The range of strange quark mass shown corresponds to the allowed range
$45< m_{\rm s}(M_Z)\!<\!135$\,MeV 
($100\!\lesssim\! m_{\rm s}(1\,\mbox{GeV})\!\lesssim\! 300$\,MeV). 
The known systematic
and statistical error is indicated by the divergence of the
different curves, and is particularly small for $N_{\rm s}/B$.

Before we can use these results to obtain a more
reliable estimate of strange quark mass, we will need to understand 
other sources of systematic uncertainty, e.g., those 
associated with unknown mechanisms of strangeness production
and to improve the model of QGP we are employing. However, 
these results are so strongly dependent on $m_{\rm s}$ and so little
on other quantities, that we can be optimistic that one day strange 
and charm flavor mass may be obtained from the flavor yields seen in 
nuclear collisions.

\begin{figure}[!t]
\vspace*{-2.5cm}
\centerline{\hspace*{-2.4cm}
\psfig{width=8.3cm,figure=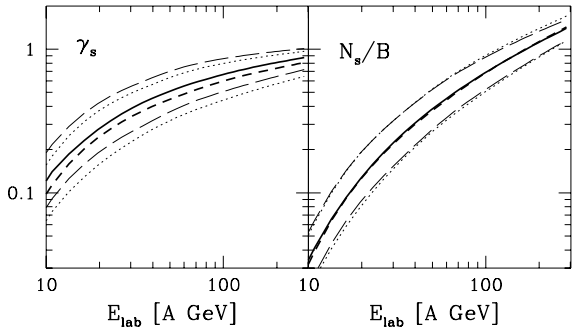}
}
\vspace*{-0.3cm}
\caption{
Phase space occupancy $\gamma_{\rm s}$ and yield of strange quarks per 
baryon $N_{\rm s}/B$ as function of laboratory collision energy in Pb--Pb 
interactions for $m_{\rm s}=200\pm40$ MeV and two different stopping 
fractions: $\eta=50\%$ (dashed and two doted lines) and $\eta=75\%$
(solid and two long dashed lines). 
\label{fignsgselab}
}
\end{figure}

Assuming that the model we proposed has been effectively
tested at 200 A GeV  S--W/Pb collisions, we
compute the strangeness yield and phase space occupancy as function
of energy. The results are shown for the case of Pb--Pb collisions. 
We show these results
in Fig.\,\ref{fignsgselab} as function of laboratory collision energy, 
stressing that if at low energies the QGP phase is not encountered, we would
expect to see a drop in strangeness yield beyond the expectations here
presented.

\begin{figure}[t]
\vspace*{-2.5cm}
\centerline{\hspace*{-2.4cm}
\psfig{width=8.3cm,figure=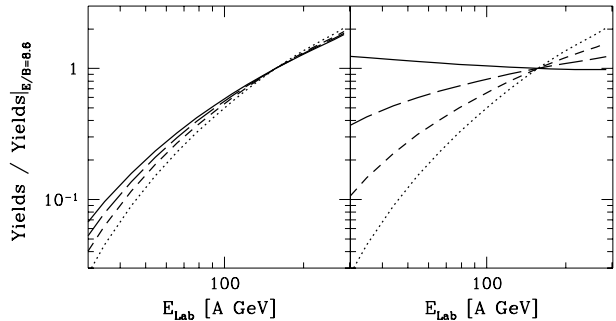}
}
\vspace*{-0.3cm}
\caption{
Relative yields of antibaryons (left) and baryons (right) as
function of heavy ion collision energy $E_{\rm Lab}$. 
Abundance set to a common value (unity) at 
$E_{\rm Lab}=158$ A GeV. Solid lines nucleons, 
long dashed $\Lambda+\Sigma^0$, short dashed $\Xi$, 
dotted $\Omega$. \label{figyields}
}
\end{figure}

These results allow us to evaluate the strange (anti)baryon yields
from QGP as function of collision energy. We note that at fixed
$m_\bot$ the medium dependent factor controlling the abundance of
hadrons emerging from the surface of the deconfined region is
related to the chemical conditions in the source, and for strange
quarks, there is also the occupancy factor $\gamma_{\rm s}$ to be
considered:
\begin{equation}
n_h|_{m_\bot} =e^{-m_\bot/T}\prod_{k\in h}\gamma_k \lambda_k\,.
\end{equation}
The strange quark fugacity is in deconfined phase unity, while the
light quark fugacity evolution with energy of colliding ions
follows from our earlier studies\,\cite{init}. In 
Fig.\,\ref{figyields}. we have normalized all yields 
at $E_{\rm Lab}=158$ A GeV. Remarkably, all antibaryon yields (left
hand side of Fig.\,\ref{figyields}) cluster together (solid lines:
(anti)nucleons, long dashed: (anti)hyperons, short-dashed:
(anti)cascades, and dotted: (anti)omegas), thus as long as the QGP
phase is formed,  ratios of rare multistrange antibaryons 
should not change significantly while the collision energy is 
reduced, until the QGP formation is disrupted. It should be noted
that the yield of $\overline{\Omega}$ remains appreciable, all the
way even at very small energies --- this is the case as long as
these particles are produced by the deconfined phase, rather 
than in individual hadronic interactions. For baryons (right hand
side of Fig.\,\ref{figyields}) there is considerable 
differentiation of the yield behavior: the reference yield of
nucleons rises slightly to compensate for the drop in the yield of 
other strange baryons, which  decreases substantially with energy. 

\section{Final Remarks}\label{endrem}
In conclusion, we can once more remind ourselves about the two 
generic strangeness observables. The relative total abundance of 
strangeness is most related to the initial condition, the `hotter' 
the initial state is, the greater the production rate, and thus the 
final state relative yield, to be measured with respect to 
baryon number or global particle multiplicity (entropy). 
The phase space occupancy of strangeness $\gamma_{\rm s}$
depends aside of the initial production rate, on the final
state dilution characterized by dynamics of the expansion and the
freeze-out temperature. Our results show that we can use certain features
of strange particle production to see the formation of
 the deconfined state and
to study some QCD properties and parameters.
 Our results suggest that already
at present energies deconfinement is attained, 
and we have explored a number of
features as function of collision energy in 
order to see if a more systematic
study is capable to confirm this conclusion.

Using QCD renormalization group methods we have studied the s and c
flavor chemical equilibrium relaxation times. We have shown that
the newly measured QCD coupling constant comprises sufficiently
small uncertainty to allow precise evaluation of strangeness
production at and below 1 GeV energy scale. Our study has further
proven that it is essential to incorporate in the evaluation of
flavor production rates both running coupling constant {\it and}
running mass. 
 
We find that running of the QCD parameters is of major 
significance, since, e.g., the effective charm production mass
is considerably reduced, seen on the scale of available 
thermal energies. We found considerable enhancement of charm
production  for temperatures applicable at SPS collision energy,
compared to fixed mass results. While charm  experiences at low
temperature $T\simeq 200$ MeV  a 100 times slower approach to
chemical equilibrium compared to strangeness, for temperatures of
about 500 MeV, as may apply to the conditions generated at LHC  or
perhaps even RHIC collider, $\tau_{\rm c}\to 30$ fm, which is
within factor two of the expected maximum lifespan of the
deconfined state. Thus our calculations suggest that there
will be a significant abundance of thermal charm in nuclear
interactions at RHIC/LHC. In consequence, open charm should play a
similar role in the diagnosis of the `hot'  $T\simeq 500$ MeV
deconfined state as strangeness is playing today for the `cold'
$T\simeq 250$ MeV case, and charm equilibrium appears within reach
of the extreme conditions possibly arising at LHC. 
 
Our here presented  results
imply that in key features  the strange particle production results
obtained at 160--200 A GeV, are consistent with the QGP formation
hypothesis. However, in order to ascertain the possibility that
indeed the QGP phase is already formed today, a more systematic 
experimental exploration as function of collision energy of the
different observable is required, for which purpose we also have
explored  the collision energy dependence of the most
characteristic strange particle features expected from the QGP
phase. 
 
\nopagebreak
\section*{Acknowledgments}
  J.R. acknowledges partial support by  DOE, grant
               DE-FG03-95ER40937\,.\\ 
{LPTHE: Unit\'e  associ\'ee au CNRS UA 280.}


\section*{References}
\begin{thebibliography}{99}
\bibitem{Sto97} R. Stock, {\it High Energy
Nuclear Interactions and Heavy Ion Collisions}, 
in this proceedings.

\bibitem{Bjo83} J.D. Bjorken, 
{\it Phys. Rev.} {\bf D27}, 140 (1983). 

\bibitem{Ody97} G. Odyniec, {\it Intriguing Results on Strangeness
Production at CERN-SPS Energies}, 
in this proceedings.

\bibitem{acta96} J. Rafelski, J. Letessier and A. Tounsi,
{\it Acta Phys. Pol.} {\bf B27}, 1035 (1996); 
and references therein.
 
\bibitem{Pet97}  P. Petiau, {\it Anomalous $J\!/\!\Psi$
Suppression in Pb--Pb Interactions at 158 A GeV/c},
in this proceedings.

\bibitem{Pfe97}  A. Pfeiffer, {\it Low-mass Electron Pair 
Production in Pb--Au Collisions at the CERN SPS --- New Results 
from the  CERES/NA45 Experiment}, 
in this proceedings.

\bibitem{Red97} K. Redlich, {\it Soft Dilepton Production in 
Ultrarelativistic Heavy Ion Collisions at SPS Energy},
in this proceedings.

\bibitem{S95} J. Rafelski, Ed., {\it Strangeness in Hadronic Matter}, 
American Institute of Physics Conference Proceedings 340, 
New York 1995.

\bibitem{sprodQGP}
T. Bir\'o and J. Zim\'anyi, {\it Phys. Lett.} {\bf B113} 
(1982) 6; {\it Nucl. Phys.} {\bf A395} (1983) 525;\\
J. Rafelski and B. M\" uller, {\it Phys. Rev. Lett.} {\bf  48} 
(1982) 1066; {\bf 56} (1986) 2334E.
 
\bibitem{impact96}  J. Letessier, J. Rafelski, and A. Tounsi,
{\it Impact of QCD and QGP properties on
strangen\-ess production,} {\it Phys. Lett.} B. {\it in press}~(1996).
 
\bibitem{Elm95} P. Elmfors and R. Kobes, {\it Phys. Rev.} {\bf D51}, 774
(1995).
 
\bibitem{SS96} L.R. Surguladze and M.A. Samuel, 
{\it Rev. Mod. Phys.} {\bf 68}  (1996) 259.
 
\bibitem{ICHEP96} M. Schmelling, {\it Status of the Strong Coupling
Constant},
in this proceedings.
 
\bibitem{init} J. Letessier, J. Rafelski, and A. Tounsi,
{\it Phys. Lett.} {\bf B333}, 484 (1994); {\bf B323}, 393 (1994).
 
\bibitem{GR96}
M. Ga\'zdzicki and D. R\"ohrich, 
{\it Strangeness in Nuclear Collisions}
preprint IKF-HENPG/8-95, {\it Z. Physik} C (1996).

\end{thebibliography}
\end{document}